\documentclass[fleqn,preprint,preprintnumbers,amsmath,amssymb]{revtex4}
\usepackage{graphicx}
\usepackage{dcolumn}
\usepackage{bm}

\begin{document}

\title{Higher-order nonlinear electron-acoustic solitary excitations in partially degenerate quantum electron-ion plasmas}
\author{M. Akbari-Moghanjoughi}
\affiliation{Azarbaijan University of
Tarbiat Moallem, Faculty of Sciences,
Department of Physics, 51745-406, Tabriz, Iran}

\date{\today \hspace{2mm}}
\begin{abstract}
Propagation of dressed solitary excitations are studied in a partially degenerate quantum plasma in the framework of quantum-hydrodynamics (QHD) model using multiple scales technique. The evolution equation together with a linear inhomogeneous differential equation is solved using Kodama-Taniuti renormalizing technique. It is shown that the type of solitary excitations (bright or dark) is defined by two critical plasma parameter values.
PACs: 52.30.Ex, 52.35.-g, 52.35.Fp, 52.35.Mw
\end{abstract}


\maketitle

\section{Introduction}

Quantum effects are of vital importance on collective phenomena in dense plasmas \cite{Manfredi} which may be encountered in astrophysical objects like white dwarfs, active galactic nuclei and neutron stars \cite{Silva} or in metals and artificially manufactured compounds like semiconductors and nano-structured materials \cite{Haug, Markowich}. One of the key features of dense plasmas is the quantum degeneracy effect due to Pauli exclusion principal which starts playing an effective role when the de Broglie thermal wavelength $\lambda_B = h/(2\pi m_e k_B T)^{1/2}$ of the plasma ingredients is larger than the average inter-particle distances, $n^{-1/3}$ \cite{Bonitz} or equivalently when the particles temperature is below the characteristic Fermi-temperature. This condition is well satisfied for metals and dense astrophysical objects. On the other hand, a zero temperature Fermi gas model, in which all electrons are in degenerated state, may be an idealistic assumption for real hot quantum plasmas such as white dwarfs \cite{shapiro}. Therefore, at some circumstances such as the white dwarf surface one may expect two types of electrons, namely, degenerate and non-degenerate.

On the other hand, recently the field of quantum plasmas, has achieved much attention due to diverse applications in rapidly growing miniaturization technology \cite{Marklund1, Shukla1, Shaikh, Brodin1, Marklund2, Brodin2, Marklund3}. Different aspects of nonlinear wave phenomenon in quantum plasma systems has been studied using quantum hydrodynamics (QHD) model \cite{Gardner, Shukla2, Hass, Sah1}. Also, higher nonlinearity effects which is supposed to play significant role in nonlinear wave dynamics in quantum plasmas has been investigated in electron-positron-ion plasmas, where it has been shown that the quantum tunneling effects, caused due to quantum force, gives rise to higher dispersion rather than dissipation of solitary propagations and greatly affects the nonlinear wave dynamics \cite{Chatterjee}.

The small amplitude electron-acoustic solitary excitations in partially degenerate (two-temperature) quantum plasmas has been studies by number of authors \cite{mahmood, sah2, masood, moslem}. Misra et.al \cite{misra} have found ranges of cold-to-hot electron density ratio and the quantum diffraction parameter values which correspond to different types of solitary excitations. Also, recent investigation of electron-acoustic solitary propagations has shown that critical hot-to-cold electron densities can exist in two electron species quantum plasmas \cite{sahu}. On the other hand, it has been noted that the propagation of the electron-acoustic waves remain undamped in the range $0.25<n_{c0}/n_{h0}<4$ \cite{tokar}, where, $n_{c0}$ and $n_{h0}$ denote the cold and hot electrons equilibrium densities, respectively.

In current study the QHD model is applied to investigate the higher-nonlinearity effects on electron-acoustic solitary excitations in two-electron-type quantum electron-ion plasma. The results obtained here cab be helpful in astrophysical dense-plasma such as white dwarfs. The organization of the article is as follows. Description of quantum hydrodynamics state equations is given in Section II. Reductive perturbation method is applied and the Korteweg de Vries (KdV) evolution equation is obtained in Section III. The stationary solution to higher-order soliton amplitude approximation is given in Section IV. Section V presents the discussions based on the numerical analysis and, finally, Section VI devotes to the concluding remarks.

\section{Quantum Hydrodynamics Model}

In this section we consider a collision-less unmagnetized quantum plasma with two types of electrons, namely, degenerate and non-degenerate and additional neutralizing background inertial ions. This model may be appropriate for characterizing a plasma near the gaseous fringe of a white dwarf \cite{Chandra}, where both degenerate and non-degenerate electron populations may be present. The degeneracy pressure only applies to hot ions with temperature below the corresponding Fermi temperature. On the other hand, the collisions in such plasma are assumed to be limited due to Fermi-blocking process, hence, to be considered as collision-less. The closed set of basic quantum hydrodynamics (QHD) equations, involving the potential tunneling force, can be written as
\begin{equation}\label{general}
\begin{array}{l}
\frac{{\partial {n_d}}}{{\partial t}} + \frac{{\partial {u_d}{n_d}}}{{\partial x}} = 0, \\
\frac{{\partial {n_n}}}{{\partial t}} + \frac{{\partial {u_n}{n_n}}}{{\partial x}} = 0, \\
\frac{{\partial {u_d}}}{{\partial t}} + {u_d}\frac{{\partial {u_d}}}{{\partial x}} =  \frac{{{Z_d}}}{{{m_d}}}\frac{{\partial \phi }}{{\partial x}} - \frac{1}{{{m_d}{n_d}}}\frac{{\partial {P_d}}}{{\partial x}} + \frac{{{\hbar ^2}}}{{2{m_d}^2}}\frac{\partial }{{\partial x}}\left[ {\frac{1}{{\sqrt {{n_d}} }}\frac{{{\partial ^2}\sqrt {{n_d}} }}{{\partial {x^2}}}} \right], \\
\frac{{\partial {u_n}}}{{\partial t}} + {u_n}\frac{{\partial {u_n}}}{{\partial x}} =  \frac{{{Z_n}}}{{{m_n}}}\frac{{\partial \phi }}{{\partial x}} + \frac{{{\hbar ^2}}}{{2{m_n}^2}}\frac{\partial }{{\partial x}}\left[ {\frac{1}{{\sqrt {{n_n}} }}\frac{{{\partial ^2}\sqrt {{n_n}} }}{{\partial {x^2}}}} \right], \\
\frac{{{\partial ^2}\phi }}{{\partial {x^2}}} = \frac{e}{{{\varepsilon _0}}}({Z_d}{n_d} + {Z_n}{n_n} - {Z_i}{N_i}). \\
\end{array}
\end{equation}
We used the subscripts $n$ and $d$ to identify the non-degenerate and degenerate plasma species, respectively. The parameter $\hbar$ is the normalized Plank constant and $N_i$ and $Z_i$ denote the background-ion density and charge. It is noted that, $Z_n=Z_d=1$ and $m_n=m_d$. From the standard definitions the degeneracy pressure relates to the degenerate electrons number-density through the following relation
\begin{equation}\label{P}
{P_d} = \frac{{{m_d}v_{Fd}^2}}{{3n_{d,0}^2}}n_d^3,\hspace{3mm}{v_{Fd}} = \sqrt {\frac{{2{E_{Fd}}}}{{{m_d}}}},\hspace{3mm}{E_{Fd}} = {k_B}{T_{Fd}},
\end{equation}
where, quantities $v_{Fd}$, $E_{Fd}$ and $T_{Fd}$ denote Fermi-velocity, Fermi-energy and Fermi-temperature, respectively and $n_{d,0}$ is the degenerate electrons equilibrium number-density. Under the zero-temperature assumption for degenerated electrons the Fermi-temperature is related to the number-density via the relation
\begin{equation}\label{T}
{T_{Fd}} = \frac{E_{Fd}}{k_B} = \frac{{{\hbar ^2}}}{{2{m_d}{k_B}}}{(3{\pi ^2}{n_{d,0}})^{3/2}}.
\end{equation}
On the other hand, we use the following scalings together with Eqs. (\ref{T}) and (\ref{P}) to obtain a dimensionless set of QHD equations, as
\begin{equation}
x \to \frac{{{C_{sd}}}}{{{\omega _{pd}}}}x,\hspace{3mm}t \to \frac{t}{{{\omega _{pd }}}},\hspace{3mm}n \to n{n_{d,0}},\hspace{3mm}u \to u{C_{sd}},\hspace{3mm}\varphi  \to \varphi \frac{{2{k_B}{T_{Fd }}}}{e}.
\end{equation}
where, ${\omega _{pd }} = \sqrt {{e^2}{n_{d,0}}/{\varepsilon _0}{m_d }}$ and ${C_{sd }} = \sqrt {2{k_B}{T_{Fd}}/{m_d }}$ are the characteristic quantum plasmon frequency and Fermi-speed, respectively.
The normalized set of equations read as
\begin{equation}\label{normal}
\begin{array}{l}
\frac{{\partial {n_d}}}{{\partial t}} + \frac{{\partial {u_d}{n_d}}}{{\partial x}} = 0, \\
\frac{{\partial {n_n}}}{{\partial t}} + \frac{{\partial {u_n}{n_n}}}{{\partial x}} = 0, \\
\frac{{\partial {u_d}}}{{\partial t}} + {u_d}\frac{{\partial {u_d}}}{{\partial x}} = \frac{{\partial \phi }}{{\partial x}} - {n_d}\frac{{\partial {n_d}}}{{\partial x}} + \frac{{{H^2}}}{2}\frac{\partial }{{\partial x}}\left[ {\frac{1}{{\sqrt {{n_d}} }}\frac{{{\partial ^2}\sqrt {{n_d}} }}{{\partial {x^2}}}} \right], \\
\frac{{\partial {u_n}}}{{\partial t}} + {u_n}\frac{{\partial {u_n}}}{{\partial x}} = \frac{{\partial \phi }}{{\partial x}} + \frac{{{H^2}}}{2}\frac{\partial }{{\partial x}}\left[ {\frac{1}{{\sqrt {{n_n}} }}\frac{{{\partial ^2}\sqrt {{n_n}} }}{{\partial {x^2}}}} \right], \\
\frac{{{\partial ^2}\phi }}{{\partial {x^2}}} = {n_d} + {n_n} - {Z_i}{N_i}. \\
\end{array}
\end{equation}
The quantities $u_d$ ($u_n$), $n_d$ ($n_n$) and $\phi$ refer to the velocity, density of degenerate (non-degenerate) species and the electrostatic potential, respectively. The normalized fractional parameter $H=\hbar \omega_{pd}/2k_B T_{Fd}$ is the quantum diffraction parameter, being the ratio of electron plasmon-energy to the Fermi-energy.

The quasi-neutrality condition is achieved via Poisson's relation at thermodynamic equilibrium state as
\begin{equation}
n_{d,0} +n_{n,0} -Z_i N_i =0,
\end{equation}
or in dimensional form as
\begin{equation}\label{neural}
1 +\beta =\delta,\hspace{3mm}\beta =\frac{n_{n,0} }{n_{d,0} },\hspace{3mm}\delta =\frac{Z_i N_i}{n_{d,0}}.
\end{equation}
The parameter $\beta$ is the fraction of non-degenerate to degenerate electron density which can be considered as a measure of degree of plasma degeneracy.

We may obtain the linear dispersion relation by a Fourier analysis of Eqs. (\ref{normal}) as
\begin{equation}\label{mode}
\frac{\beta }{{\frac{{{H^2}{k^2}}}{4} - \frac{{{\omega ^2}}}{{{k^2}}}}} + \frac{1}{{1 + \frac{{{H^2}{k^2}}}{4} - \frac{{{\omega ^2}}}{{{k^2}}}}} =  - {k^2}.
\end{equation}
There are higher and lower small-$k$ limits branches in dispersion relation in the following forms
\begin{equation}\label{limits}
\begin{array}{l}
{\omega ^2}_{+}  \approx 1 + \beta  + \frac{1}{{(1 + \beta )}}{k^2}+ \left[ {\frac{{{H^2}}}{4} + \frac{\beta }{{{{(1 + \beta )}^3}}}} \right]{k^4},\\
{\omega ^2}_{-}  \approx \frac{\beta }{{1 + \beta }}{k^2} + \left[ {\frac{{{H^2}}}{4} - \frac{\beta }{{{{(1 + \beta )}^3}}}} \right]{k^4}. \\
\end{array}
\end{equation}
It should be noted that the QHD model does not apply to large-$k$ values (e.g. see Ref. \cite{Hass}). The variations of dispersion curves with respect to the quantum parameter, $H$, and the degeneracy parameter, $\beta$, is given in Fig. (1). It is clearly observed that the increase in the value of quantum diffraction, $H$, causes a larger dispersion of both higher and lower modes. It is also remarked that the higher mode starts at higher frequencies as the plasma becomes less degenerate (i.e. as $\beta$ increases). On the other hand, at the special case of a classical non-degenerate plasma, $\beta\rightarrow\infty,H=0$ the dispersion relation for acoustic mode reduces to $\omega_{-} \approx k$.

\section{First-Order Amplitude Evolution}

Here, we introduce a strained coordinate in which the amplitude perturbation moves with a phase velocity of $\lambda$,
\begin{subequations}
\begin{equation}\label{stretch}
\xi =\varepsilon ^{\frac{1}{2}}\left( {x-\lambda t} \right),
\end{equation}
\begin{equation}
\tau =\varepsilon ^{\frac{3}{2}}t
\end{equation}
\end{subequations}
Hence, the normalized QHD equations can be written in the following forms in the new stretched coordinate
\begin{subequations}\label{stretchedEq}
\begin{equation}
\begin{array}{l}
\varepsilon \frac{{\partial {n_d}}}{{\partial \tau }} - \lambda \frac{{\partial {n_d}}}{{\partial \xi }} + \frac{{\partial {u_d}{n_d}}}{{\partial \xi }} = 0, \\ \varepsilon \frac{{\partial {n_n}}}{{\partial \tau }} - \lambda \frac{{\partial {n_n}}}{{\partial \xi }} + \frac{{\partial {u_n}{n_n}}}{{\partial \xi }} = 0, \\
\end{array}
\end{equation}
\begin{equation}
\begin{array}{l}
\varepsilon {n_d}^3\frac{{\partial {u_d}}}{{\partial \tau }} - \lambda {n_d}^3\frac{{\partial {u_d}}}{{\partial \xi }} + {n_d}^3{u_d}\frac{{\partial {u_d}}}{{\partial \xi }} - {n_d}^3\frac{{\partial \phi }}{{\partial \xi }} + {n_d}^4\frac{{\partial {n_d}}}{{\partial \xi }} -  \\
\varepsilon \frac{{{H^2}}}{4}\left[ {{{\left( {\frac{{\partial {n_d}}}{{\partial \xi }}} \right)}^3} - 2{n_d}\frac{{\partial {n_d}}}{{\partial \xi }}\frac{{{\partial ^2}{n_d}}}{{\partial {\xi ^2}}} + {n_d}^2\frac{{{\partial ^3}{n_d}}}{{\partial {\xi ^3}}}} \right] = 0, \\
\varepsilon {n_n}^3\frac{{\partial {u_n}}}{{\partial \tau }} - \lambda {n_n}^3\frac{{\partial {u_n}}}{{\partial \xi }} + {n_n}^3{u_n}\frac{{\partial {u_n}}}{{\partial \xi }} - {n_n}^3\frac{{\partial \phi }}{{\partial \xi }} -  \\ \varepsilon \frac{{{H^2}}}{4}\left[ {{{\left( {\frac{{\partial {n_n}}}{{\partial \xi }}} \right)}^3} - 2{n_n}\frac{{\partial {n_n}}}{{\partial \xi }}\frac{{{\partial ^2}{n_n}}}{{\partial {\xi ^2}}} + {n_n}^2\frac{{{\partial ^3}{n_n}}}{{\partial {\xi ^3}}}} \right] = 0, \\
\end{array}
\end{equation}
\begin{equation}
\frac{{{\partial ^2}\phi }}{{\partial {x^2}}} = {n_d} + {n_n} - {Z_c}{N_c},
\end{equation}
\end{subequations}
with the parameter $\varepsilon$ characterized as small, positive and real number proportional to the perturbation amplitude. The asymptotic expansion of dependent plasma variables around thermodynamics equilibrium is, then, achieved using the following expansions
\begin{subequations}\label{ordering}
\begin{equation}
\begin{array}{l}
{n_d} = 1 + \varepsilon n_d^{(1)} + {\varepsilon ^2}n_d^{(2)} +  \ldots , \\
{n_n} = \beta  + \varepsilon n_n^{(1)} + {\varepsilon ^2}n_n^{(2)} +  \ldots , \\
{u_d} = \varepsilon u_d^{(1)} + {\varepsilon ^2}u_d^{(2)} + {\varepsilon ^3}u_d^{(3)} +  \ldots , \\
{u_n} = \varepsilon u_n^{(1)} + {\varepsilon ^2}u_n^{(2)} + {\varepsilon ^3}u_n^{(3)} +  \ldots , \\
\phi =\varepsilon \phi ^{(1)}+\varepsilon ^2\phi ^{(2)}+\varepsilon
^3\phi ^{(3)}+\ldots .
\end{array}
\end{equation}
\end{subequations}
By isolation of the distinct perturbation orders, from the leading-orders in $\varepsilon$, we deduce the following relations among first-order perturbed plasma quantities
\begin{subequations}\label{firstorder}
\begin{equation}
\begin{array}{l}
\lambda \frac{{\partial n_d^{(1)}}}{{\partial \xi }} + \frac{{\partial u_d^{(1)}}}{{\partial \xi }} = 0, \\
\lambda \frac{{\partial n_n^{(1)}}}{{\partial \xi }} + \beta \frac{{\partial u_n^{(1)}}}{{\partial \xi }} = 0, \\
\end{array}
\end{equation}
\begin{equation}
\begin{array}{l}
\lambda \frac{{\partial u_d^{\left( 1 \right)}}}{{\partial \xi }} - \frac{{\partial {\phi ^{(1)}}}}{{\partial \xi }} + n_d^{\left( 1 \right)}\frac{{\partial n_d^{\left( 1 \right)}}}{{\partial \xi }} = 0, \\
\lambda \frac{{\partial u_n^{\left( 1 \right)}}}{{\partial \xi }} - \frac{{\partial \phi _n^{\left( 1 \right)}}}{{\partial \xi }} + n_n^{\left( 1 \right)}\frac{{\partial n_n^{\left( 1 \right)}}}{{\partial \xi }} = 0, \\
\end{array}
\end{equation}
\begin{equation}
n_d^{(1)} + n_n^{(1)} = 0.
\end{equation}
\end{subequations}
Therefore, the first-order approximations for degenerate and non-degenerate ion velocities and densities are found to be
\begin{subequations}\label{firstcomp}
\begin{equation}
\begin{array}{l}
u_d^{(1)} = {U_{1d}}{\phi ^{(1)}},\hspace{3mm}{U_{1d}} = \frac{\lambda }{{1 - {\lambda ^2}}} \\
u_n^{(1)} = {U_{1n}}{\phi ^{(1)}},\hspace{3mm}{U_{1n}} =  - \frac{1}{\lambda }, \\
n_d^{(1)} = {N_{1d}}{\phi ^{(1)}},\hspace{3mm}{N_{1d}} = \frac{1}{{1 - {\lambda ^2}}}, \\
n_n^{(1)} = {N_{1n}}{\phi ^{(1)}},\hspace{3mm}{N_{1n}} =  - \frac{\beta }{{{\lambda ^2}}}. \\
\end{array}
\end{equation}
\end{subequations}
The nonlinear dispersion relation as well as the phase velocity of wave, which are obtained through the compatibility requirement, read as
\begin{equation}\label{D}
\frac{1}{{1 - {\lambda ^2}}} - \frac{\beta }{\lambda } = 0,
\end{equation}
\begin{equation}
\lambda  = \sqrt {\frac{\beta }{{1 + \beta }}}.
\end{equation}
In the next higher-order of $\varepsilon$ we get
\begin{subequations}\label{secondorder}
\begin{equation}
\begin{array}{l}
\frac{{\partial n_d^{\left( 1 \right)}}}{{\partial \tau }} - \lambda \frac{{\partial n_d^{\left( 2 \right)}}}{{\partial \xi }} + \frac{{\partial u_d^{\left( 2 \right)}}}{{\partial \xi }} + \frac{{\partial u_d^{\left( 1 \right)}n_d^{\left( 1 \right)}}}{{\partial \xi }} = 0, \\
\frac{{\partial n_n^{\left( 1 \right)}}}{{\partial \tau }} - \lambda \frac{{\partial n_n^{\left( 2 \right)}}}{{\partial \xi }} + \beta \frac{{\partial u_n^{\left( 2 \right)}}}{{\partial \xi }} + \frac{{\partial u_n^{\left( 1 \right)}n_n^{\left( 1 \right)}}}{{\partial \xi }} = 0, \\
\end{array}
\end{equation}
\begin{equation}
\begin{array}{l}
\frac{{\partial u_d^{(1)}}}{{\partial \tau }} - \lambda \frac{{\partial u_d^{(2)}}}{{\partial \xi }} - 3\lambda n_d^{\left( 1 \right)}\frac{{\partial u_d^{(1)}}}{{\partial \xi }} + u_d^{(1)}\frac{{\partial u_d^{(1)}}}{{\partial \xi }} -  \\
3n_d^{\left( 1 \right)}\frac{{\partial \phi _d^{(1)}}}{{\partial \xi }} - \frac{{\partial \phi _d^{(2)}}}{{\partial \xi }} + 4n_d^{\left( 1 \right)}\frac{{\partial n_d^{\left( 1 \right)}}}{{\partial \xi }} + \frac{{\partial n_d^{\left( 2 \right)}}}{{\partial \xi }} - \frac{{{H^2}}}{4}\frac{{{\partial ^3}n_d^{\left( 1 \right)}}}{{\partial {\xi ^3}}} = 0, \\
\beta \frac{{\partial u_n^{(1)}}}{{\partial \tau }} - \lambda \beta \frac{{\partial u_n^{(2)}}}{{\partial \xi }} - 3\lambda n_n^{\left( 1 \right)}\frac{{\partial u_n^{(1)}}}{{\partial \xi }} + \beta u_n^{(1)}\frac{{\partial u_n^{(1)}}}{{\partial \xi }} -  \\
3n_n^{\left( 1 \right)}\frac{{\partial {\phi ^{(1)}}}}{{\partial \xi }} - \beta \frac{{\partial {\phi ^{(2)}}}}{{\partial \xi }} + {\beta ^2}\frac{{\partial n_n^{\left( 2 \right)}}}{{\partial \xi }} - \frac{{{H^2}}}{4}\frac{{{\partial ^3}n_n^{\left( 1 \right)}}}{{\partial {\xi ^3}}} = 0,\\
 \end{array}
\end{equation}
\begin{equation}
\frac{{{\partial ^2}{\phi ^{\left( 1 \right)}}}}{{\partial {\xi ^2}}} = n_d^{\left( 2 \right)} + n_n^{\left( 2 \right)}.
\end{equation}
\end{subequations}
Consequently, using the dispersion relation (Eq. \ref{D}) to eliminate the common terms in Eqs. (\ref{secondorder}), a KdV-type evolution equation is obtained
\begin{equation}\label{KdV}
\frac{\partial \phi ^{\left( 1 \right)}}{\partial \tau }+AB\phi ^{\left( 1 \right)}\frac{\partial
\phi ^{\left( 1 \right)}}{\partial \xi }+\frac{A}{2}\frac{\partial
^3\phi ^{\left( 1 \right)}}{\partial \xi ^3}=0.
\end{equation}
The coefficients of KdV evolution equation read as
\begin{equation}\label{A}
A =\frac{{ 4\beta - {H^2}{{(1 + \beta )}^3}}}{{8{{(1 + \beta )}^{5/2}}\sqrt \beta  }},
\end{equation}
\begin{equation}\label{B}
B = \frac{{4\beta  - 3}}{2\lambda A}.
\end{equation}
This is the equation which describes the first-order evolution of the electrostatic potential, however, in order to achieve better approximations we need to consider the higher terms in $\varepsilon$, which will follow next. First, we note that there is critical plasma value, $H_{cr}$ at which the dispersion coefficient vanishes. Also, for the value of $\beta_{cr}$ the nonlinear coefficient vanishes. Therefore, no KdV-type solitary excitations are expected for these limiting cases. To obtain a solitary solution for these limiting cases one has to solve an MKdV-type equation. These critical plasma values are given by
\begin{equation}\label{B}
\begin{array}{l}
{\beta _{cr}} = \frac{3}{4}, \\
{H_{cr}} = \frac{2}{{1 + \beta }}\sqrt {\frac{\beta }{{1 + \beta }}}  \\
\end{array}
\end{equation}
The higher-order perturbed components for degenerate and non-degenerate species can be obtained by solving Eqs. (\ref{firstcomp}) and (\ref{secondorder}) together. These solutions are of the following shapes in terms of the first and second-order potentials for degenerate electrons
\begin{equation}\label{n2}
\begin{array}{l}
{n_d}^{\left( 2 \right)} = {N_{1d}}{\phi ^{(2)}} + {N_{12d}}{\phi ^{\left( 1 \right)}}^2 + {N_{13d}}\frac{{{\partial ^2}{\phi ^{(1)}}}}{{\partial {\xi ^2}}}, \\
{u_d}^{\left( 2 \right)} = {U_{1d}}{\phi ^{\left( 2 \right)}} + {U_{12d}}{\phi ^{\left( 1 \right)}}^2 + {U_{13d}}\frac{{{\partial ^2}{\phi ^{\left( 1 \right)}}}}{{\partial {\xi ^2}}}, \\
\end{array}
\end{equation}
and, also for non-degenerate electrons
\begin{equation}\label{u2}
\begin{array}{l}
{n_n}^{\left( 2 \right)} = {N_{1n}}{\phi ^{(2)}} + {N_{12n}}{\phi ^{\left( 1 \right)}}^2 + {N_{13n}}\frac{{{\partial ^2}{\phi ^{(1)}}}}{{\partial {\xi ^2}}}, \\
{u_n}^{\left( 2 \right)} = {U_{1n}}{\phi ^{\left( 2 \right)}} + {U_{12n}}{\phi ^{\left( 1 \right)}}^2 + {U_{13n}}\frac{{{\partial ^2}{\phi ^{\left( 1 \right)}}}}{{\partial {\xi ^2}}}, \\
\end{array}
\end{equation}
with the coefficients given bellow
\begin{equation}\label{n2coeffs}
\begin{array}{l}
{N_{12d}} = \frac{1}{2}\left[ {\frac{{4{N_{1d}}^2 - 2\lambda AB{N_{1d}} - 3{N_{1d}}}}{{{\lambda ^2} - 1}}} \right], \\
{N_{13d}} =  - \frac{{{N_{1d}}}}{4}\left[ {\frac{{4\lambda A + {H^2}}}{{{\lambda ^2} - 1}}} \right], \\
{U_{12d}} = \lambda {N_{12d}} - {N_{1d}}{U_{1d}} + \frac{{AB{N_{1d}}}}{{({\lambda ^2} - 1)}}, \\
{U_{13d}} = \lambda {N_{13d}} + \frac{{A{N_{1d}}}}{2} \\
\end{array}
\end{equation}
and
\begin{equation}\label{u2coeffs}
\begin{array}{l}
{N_{12n}} = \frac{1}{2}\left[ {\frac{{4\beta {N_{1n}}^2 - 2\lambda AB{N_{1n}} - 3{N_{1n}}}}{{{\lambda ^2}}}} \right], \\
{N_{13n}} =  - \frac{{{N_{1n}}}}{4}\left[ {\frac{{4\lambda A + {H^2}}}{{{\lambda ^2}}}} \right], \\
{U_{12n}} = \frac{{\lambda {N_{12n}}}}{\beta } - \frac{{{N_{1n}}{U_{1n}}}}{\beta } + \frac{{AB{N_{1n}}\beta }}{{{\lambda ^2}}}, \\
{U_{13n}} = \frac{{\lambda {N_{13n}}}}{\beta } + \frac{{A{N_{1n}}}}{{2\beta }}. \\
\end{array}
\end{equation}
Finally, by considering the third order of approximation we obtain the next-order defining relations for both degenerate and non-degenerate electrons as
\begin{subequations}\label{thirdorder}
\begin{equation}
\begin{array}{l}
\frac{{\partial n_d^{\left( 2 \right)}}}{{\partial \tau }} - \lambda \frac{{\partial n_d^{\left( 3 \right)}}}{{\partial \xi }} + \frac{{\partial u_d^{\left( 3 \right)}}}{{\partial \xi }} + \frac{{\partial u_d^{\left( 1 \right)}n_d^{\left( 2 \right)}}}{{\partial \xi }} + \frac{{\partial u_d^{\left( 2 \right)}n_d^{\left( 1 \right)}}}{{\partial \xi }} = 0, \\
\frac{{\partial n_n^{\left( 2 \right)}}}{{\partial \tau }} - \lambda \frac{{\partial n_n^{\left( 3 \right)}}}{{\partial \xi }} + \beta \frac{{\partial u_n^{\left( 3 \right)}}}{{\partial \xi }} + \frac{{\partial u_n^{\left( 1 \right)}n_n^{\left( 2 \right)}}}{{\partial \xi }} + \frac{{\partial u_n^{\left( 2 \right)}n_n^{\left( 1 \right)}}}{{\partial \xi }} = 0, \\
\end{array}
\end{equation}
\begin{equation}
\begin{array}{l}
\frac{{\partial u_d^{\left( 2 \right)}}}{{\partial \tau }} + 3n_d^{\left( 1 \right)}\frac{{\partial u_d^{(1)}}}{{\partial \tau }} - \lambda \frac{{\partial u_d^{\left( 3 \right)}}}{{\partial \xi }} - 3\lambda n_d^{\left( 1 \right)}\frac{{\partial u_d^{\left( 2 \right)}}}{{\partial \xi }} - 3\lambda {(n_d^{\left( 1 \right)})^2}\frac{{\partial u_d^{\left( 1 \right)}}}{{\partial \xi }} - 3\lambda n_d^{\left( 2 \right)}\frac{{\partial u_d^{\left( 1 \right)}}}{{\partial \xi }} +  \\
u_d^{(1)}\frac{{\partial u_d^{(2)}}}{{\partial \xi }} + 3u_d^{(1)}n_d^{\left( 1 \right)}\frac{{\partial u_d^{(1)}}}{{\partial \xi }} - \frac{{\partial {\phi ^{(3)}}}}{{\partial \xi }} - 3n_d^{\left( 1 \right)}\frac{{\partial {\phi ^{(2)}}}}{{\partial \xi }} - 3{(n_d^{\left( 1 \right)})^2}\frac{{\partial {\phi ^{(1)}}}}{{\partial \xi }} -  \\
3n_d^{\left( 2 \right)}\frac{{\partial {\phi ^{(1)}}}}{{\partial \xi }} + \frac{{\partial n_d^{\left( 3 \right)}}}{{\partial \xi }} + 4n_d^{\left( 1 \right)}\frac{{\partial n_d ^{\left( 2 \right)}}}{{\partial \xi }} + 6{(n_d^{\left( 1 \right)})^2}\frac{{\partial n_d^{\left( 1 \right)}}}{{\partial \xi }} + 4n_d^{\left( 2 \right)}\frac{{\partial n_d^{\left( 1 \right)}}}{{\partial \xi }} -  \\
\frac{{{H^2}}}{2}\frac{{\partial n_d^{\left( 1 \right)}}}{{\partial \xi }}\frac{{{\partial ^2}n_d^{\left( 1 \right)}}}{{\partial {\xi ^2}}} - \frac{{{H^2}}}{4}\frac{{{\partial ^3}n_d^{\left( 2 \right)}}}{{\partial {\xi ^3}}} - \frac{{{H^2}}}{2}n_d^{\left( 1 \right)}\frac{{{\partial ^3}n_d^{\left( 1 \right)}}}{{\partial {\xi ^3}}} = 0, \\
{\beta ^2}\frac{{\partial u_n^{\left( 2 \right)}}}{{\partial \tau }} + 3{\beta}n_n^{\left( 1 \right)}\frac{{\partial u_n^{(1)}}}{{\partial \tau }} - \lambda {\beta ^3}\frac{{\partial u_n^{\left( 3 \right)}}}{{\partial \xi }} - 3\lambda {\beta}n_n^{\left( 1 \right)}\frac{{\partial u_n^{\left( 2 \right)}}}{{\partial \xi }} - 3\lambda{(n_n^{\left( 1 \right)})^2}\frac{{\partial u_n^{\left( 1 \right)}}}{{\partial \xi }} -  \\
3\lambda {\beta}n_n^{\left( 2 \right)}\frac{{\partial u_n^{\left( 1 \right)}}}{{\partial \xi }} + {\beta ^2}u_n^{(1)}\frac{{\partial u_n^{(2)}}}{{\partial \xi }} + 3{\beta ^2}u_n^{(1)}n_n^{\left( 1 \right)}\frac{{\partial u_n^{(1)}}}{{\partial \xi }} - {\beta ^3}\frac{{\partial {\phi ^{(3)}}}}{{\partial \xi }} -  \\
3{\beta}n_n^{\left( 1 \right)}\frac{{\partial {\phi ^{(2)}}}}{{\partial \xi }} - 3{(n_n^{\left( 1 \right)})^2}\frac{{\partial {\phi ^{(1)}}}}{{\partial \xi }} - 3{\beta}n_n^{\left( 2 \right)}\frac{{\partial {\phi ^{(1)}}}}{{\partial \xi }} - \frac{{{H^2}}}{2}\frac{{\partial n_n^{\left( 1 \right)}}}{{\partial \xi }}\frac{{{\partial ^2}n_n^{\left( 1 \right)}}}{{\partial {\xi ^2}}} -  \\
\frac{{{H^2}}}{4}{\beta}\frac{{{\partial ^3}n_n^{\left( 2 \right)}}}{{\partial {\xi ^3}}} - \frac{{{H^2}}}{2}n_n^{\left( 1 \right)}\frac{{{\partial ^3}n_n^{\left( 1 \right)}}}{{\partial {\xi ^3}}} = 0, \\
\end{array}
\end{equation}
\begin{equation}
\frac{{{\partial ^2}{\phi ^{\left( 2 \right)}}}}{{\partial {\xi ^2}}} = n_d^{\left( 3 \right)} + n_n^{\left( 3 \right)}.
\end{equation}
\end{subequations}
Equations (\ref{thirdorder}) together with Eqs. (\ref{firstcomp}), (\ref{KdV}), (\ref{n2coeffs}) and (\ref{u2coeffs}) give rise to the following linear inhomogeneous equation, the solution of which along with Eq. (\ref{KdV}), describes the second-order nonlinear electron-acoustic wave amplitude evolution
\begin{equation}\label{seconKdV}
\begin{array}{l}
\frac{{\partial {\phi ^{\left( 2 \right)}}}}{{\partial \tau }} + AB\frac{{\partial {\phi ^{\left( 1 \right)}}{\phi ^{\left( 2 \right)}}}}{{\partial \xi }} + \frac{A}{2}\frac{{{\partial ^3}{\phi ^{\left( 2 \right)}}}}{{\partial {\xi ^3}}} =  \\
D_0\left[ {{D_1}{{\frac{{\partial {\phi ^{\left( 1 \right)}}}}{{\partial \tau }}}^2} + {D_2}\frac{{\partial {\phi ^{(1)}}}}{{\partial \xi }}\frac{{{\partial ^2}{\phi ^{(1)}}}}{{\partial {\xi ^2}}} + {D_3}{\phi ^{\left( 1 \right)}}\frac{{{\partial ^3}{\phi ^{(1)}}}}{{\partial {\xi ^3}}} + {D_4}{\phi ^{(1)}}^2\frac{{\partial {\phi ^{(1)}}}}{{\partial \xi }} + {D_5}\frac{{{\partial ^5}{\phi ^{(1)}}}}{{\partial {\xi ^5}}}} \right]. \\
\end{array}
\end{equation}
The unknown coefficients in the linear inhomogeneous equation are given as
\begin{equation}\label{Ds}
\begin{array}{l}
{D_0} = \frac{{\sqrt \beta  }}{{2{{(1 + \beta )}^{5/2}}}}, \\
{D_1} = \frac{{(3 + \beta (3\beta  - 1)){{(1 + \beta )}^{5/2}}}}{{2{\beta ^{3/2}}}}, \\
{D_2} = - \frac{{(1 + \beta )({H^2}{{(1 + \beta )}^3}(1 + 50\beta ) + 12\beta (3 + 2\beta (4\beta  - 9)))}}{{16{\beta ^2}}},  \\
{D_3} = - \frac{{(1 + \beta )(4(7 - 22\beta )\beta  + {H^2}{{(1 + \beta )}^3}(3 + 14\beta ))}}{{16{\beta ^2}}},  \\
{D_4} = \frac{{{{(1 + \beta )}^2}(\beta (\beta (\beta (149 + 100\beta ) - 36) - 97) - 36)}}{{4{\beta ^2}}}, \\
{D_5} = \frac{{8{H^2}\beta {{(1 + \beta )}^3} + {H^4}{{(1 + \beta )}^6} + 16{\beta ^2}(4\beta  - 3)}}{{64{\beta ^2}{{(1 + \beta )}^2}}}. \\
\end{array}
\end{equation}

\section{Evolution of Dressed Solitary Excitations}

In order to obtain the second-order approximation to dressed amplitude evolution, the Eqs. (\ref{KdV}) and (\ref{seconKdV}) have to be solved together in a bounded form. To obtain such solution, we will employ the renormalizing technique introduced by Kodama and Taniuti \cite{Kodama}. Therefore, we first rewrite the Eq. (\ref{seconKdV}) in the following form
\begin{equation}\label{skdv}
\frac{{\partial {\phi ^{\left( 2 \right)}}}}{{\partial \tau }} + AB\frac{{\partial {\phi ^{\left( 1 \right)}}{\phi ^{\left( 2 \right)}}}}{{\partial \xi }} + \frac{A}{2}\frac{{{\partial ^3}{\phi ^{\left( 2 \right)}}}}{{\partial {\xi ^3}}} = \Gamma({\phi ^{\left( 1 \right)}})
\end{equation}
where,
\begin{equation}\label{s}
\Gamma\left( {{\phi ^{(1)}}} \right) = D_0\left[ {{D_1}{{\frac{{\partial {\phi ^{\left( 1 \right)}}}}{{\partial \tau }}}^2} + {D_2}\frac{{\partial {\phi ^{\left( 1 \right)}}}}{{\partial \xi }}\frac{{{\partial ^2}{\phi ^{\left( 1 \right)}}}}{{\partial {\xi ^2}}} + {D_3}{\phi ^{\left( 1 \right)}}\frac{{{\partial ^3}{\phi ^{\left( 1 \right)}}}}{{\partial {\xi ^3}}} + {D_4}{\phi ^{(1)}}^2\frac{{\partial {\phi ^{(1)}}}}{{\partial \xi }} + {D_5}\frac{{{\partial ^5}{\phi ^{(1)}}}}{{\partial {\xi ^5} }}} \right].
\end{equation}
Note that $\tilde\phi$ has be introduced for the renormalized amplitude component. For the equation set describing the first and second-order amplitude evolution, we have
\begin{subequations}\label{mkdv}
\begin{equation}
\frac{\partial \tilde\phi ^{(1)}}{\partial \tau }+AB\tilde\phi
^{(1)}\frac{\partial \tilde\phi ^{(1)}}{\partial \xi
}+\frac{A}{2}\frac{\partial ^3\tilde\phi ^{(1)}}{\partial \xi ^3}+\delta
M\frac{\partial \tilde\phi ^{(1)}}{\partial \xi }=0,
\end{equation}
\begin{equation}\label{msecondkdv}
\frac{\partial \tilde\phi ^{\left( 2 \right)}}{\partial \tau }+AB\frac{\partial
\tilde\phi ^{\left( 1 \right)}\tilde\phi ^{\left( 2 \right)}}{\partial \xi
}+\frac{A}{2}\frac{\partial ^3\tilde\phi ^{\left( 2 \right)}}{\partial \xi
^3}+\delta M\frac{\partial \tilde\phi ^{(2)}}{\partial \xi }=\Gamma\left(
{\tilde\phi ^{(1)}} \right)+\delta M\frac{\partial \tilde\phi
^{(1)}}{\partial \xi }.
\end{equation}
\end{subequations}
Following the methodology, also described in Ref. \cite{Esfand}, a moving frame $\zeta =\xi -\left( {M+\delta M} \right)\tau$ is introduced, in which Eqs. (\ref{mkdv}) read as
\begin{subequations}\label{lastkdv}
\begin{equation}\label{lkdv}
\frac{\partial ^2\tilde\phi ^{(1)}}{\partial \zeta ^2}+\left( {B\tilde\phi
^{(1)}-\frac{2M}{A}} \right)\tilde\phi ^{(1)}=0,
\end{equation}
\begin{equation}
\frac{\partial ^2\tilde\phi ^{\left( 2 \right)}}{\partial \zeta ^2}+2\left(
{B\tilde\phi ^{\left( 1 \right)}-\frac{M}{A}} \right)\tilde\phi ^{\left( 2
\right)}=\frac{2}{A}\left[\int_{-\infty }^\zeta { {\Gamma\left( {\tilde\phi ^{(1)}}
\right)} d\zeta } +\delta M\tilde\phi ^{(1)}\right],
\end{equation}
\end{subequations}
However, the desired bounded single-soliton stationary solutions should satisfy the following boundary conditions
\begin{equation}\label{boundary}
\mathop {\lim }\limits_{\zeta \to \infty } \{\tilde\phi ^{(1)},\tilde\phi
^{(2)},\frac{\partial \tilde\phi ^{(1)}}{\partial \zeta },\frac{\partial \tilde\phi
^{(2)}}{\partial \zeta },\frac{\partial ^2\tilde\phi ^{(1)}}{\partial \zeta
^2},\frac{\partial ^2\tilde\phi ^{(2)}}{\partial \zeta ^2}\}=0.
\end{equation}
We simply obtain a KdV-type solution for the first-order amplitude from Eq. (\ref{lkdv}) as
\begin{equation}\label{phi-1}
{\tilde\phi ^{(1)}} = {\tilde\phi _0}\text{sech}^2\left( {\frac{\zeta }{\Lambda }} \right),\hspace{3mm}\Lambda  = \sqrt {\frac{{2A}}{M}},\hspace{3mm}{\tilde\phi _0} = \frac{{3M}}{{AB}}.
\end{equation}
Now, using Eq. (\ref{s}) and the solution Eq. (\ref{phi-1}) the right hand side of Eq. (\ref{lastkdv}) is rewritten as
\begin{equation}\label{ms}
\begin{array}{l}
\int_{ - \infty }^\zeta  {\left[ {\Gamma \left( {{{\tilde \phi }^{\left( 1 \right)}}} \right)} \right]d\zeta  + \delta M {{\tilde \phi }^{\left( 1 \right)}}}  =  \\ \frac{{{D_0}{{\tilde \phi }_0}}}{{3{\Lambda ^4}}}\left[ {360{D_5} + {\Lambda ^2}{{\tilde \phi }_0}\left[ {(12A{D_1} - 6{D_2} + 12{D_3}) + ({D_4} - 2AB{D_1}){\Lambda ^2}{{\tilde \phi }_0}} \right]} \right]{\rm{sec}}{{\rm{h}}^6}\left( {\frac{\zeta }{\Lambda }} \right) -\\ \frac{{{D_0}{{\tilde \phi }_0}}}{{{\Lambda ^4}}} \left[ {120{D_5} - 2{\Lambda ^2}{{\tilde \phi }_0}({D_2} + {D_3} + A{D_1})} \right]{\rm{sec}}{{\rm{h}}^4}\left( {\frac{\zeta }{\Lambda }} \right) + {\tilde \phi _0}\left( {\frac{{16{D_0}{D_5}}}{{{\Lambda ^4}}} + \delta M } \right){\rm{sec}}{{\rm{h}}^2}\left( {\frac{\zeta }{\Lambda }} \right)
\end{array}
\end{equation}
The last term in Eq. (\ref{ms}) is set equal to zero to eliminate the secular term in Eq. (\ref{lastkdv}) which rather simplifies Eq. (\ref{lastkdv}) to the following form
\begin{equation}\label{fsecondkdv}
\begin{array}{l}
\frac{{{\partial ^2}{\tilde\phi ^{\left( 2 \right)}}}}{{\partial {\zeta ^2}}} + 2\left(B\tilde\phi_0{\text{sech}^2}\left( {\frac{\zeta }{\Lambda }} \right) - \frac{M}{A} \right){\tilde\phi ^{\left( 2 \right)}} =  {C_1}{\rm{sec}}{{\rm{h}}^6}\left( {\frac{\zeta }{\Lambda }} \right) -
{C_2}{\rm{sec}}{{\rm{h}}^4}\left( {\frac{\zeta }{\Lambda }} \right), \\
\end{array}
\end{equation}
where the $C$ coefficients are given as
\begin{equation}\label{ki}
\begin{array}{l}
{C_1} = \frac{{2{D_0}{\tilde\phi _0}}}{{3A{\Lambda ^4}}}\left[ {360{D_5} + {\Lambda ^2}{{\tilde \phi }_0}\left[ {(12A{D_1} - 6{D_2} + 12{D_3}) + ({D_4} - 2AB{L_1}){\Lambda ^2}{{\tilde \phi }_0}} \right]} \right],\\
{C_2} = \frac{{2{D_0}{{\tilde \phi }_0}}}{{A{\Lambda ^4}}}\left[ {120{D_5} - 2{\Lambda ^2}{{\tilde \phi }_0}({D_2} + {D_3} + A{D_1})} \right].
\end{array}
\end{equation}
Employing the "$\tanh$" method i.e., by changing the variable $\eta = \tanh (\frac{\zeta }{\Lambda})$, the Eq. (\ref{fsecondkdv}) is transformed into the familiar form of associated Legendre function \cite{Esfand}. Hence, the total amplitude of the perturbed electrostatic potential up to the second-order approximation, after absorbing the $\varepsilon$ and $\varepsilon^2$ coefficients, can be written as
\begin{equation}\label{final}
\tilde \phi_{tot} \left( \zeta  \right) = {\tilde \phi ^{\left( 1 \right)}} + {\tilde \phi ^{\left( 2 \right)}} = \text{sech}^2\left( {\frac{\zeta }{\Lambda }} \right)\left[ {{{\tilde \phi }_0} +\left( {\frac{{{C_2}}}{6} + \frac{{{C_1}}}{4} - \frac{{{C_1}}}{8}\text{sech}^2\left( {\frac{\zeta }{\Lambda }} \right)} \right)} \right].
\end{equation}

\section{Numerical Evaluation And Discussion}

Evaluation of the first-order amplitude solution reveals that the soliton amplitude diverges at the critical value of $\beta_{cr}=3/4$. In this analysis we use only values of $\beta\ll 1$ ($n_{0,n}\ll n_{0,d}$) where the $\tilde \phi_0 \ll 1$ is satisfied. Figure 2(a) shows two distinct regions (separated by the critical $H$-curve) where the solitary bright- or dark-type excitations can exist in this plasma. For the upper/lower region only bright/dark solitons exist. It is also shown in Fig. 2(b) that, depending on the value of matching-speed $M=\pm 0.1$, the hight/depth (first-order amplitude) of compressive/rarefactive solitary profile increases as the fraction of equilibrium nondegenerate to degenerate electron density increases. The distinct branches (interconnected at a value of $H_{cr}$) shown in Fig. 2(c) correspond to compressive (left-branch) and rarefactive (right-branch) solitary wave and indicate that the soliton width vanishes at $H_{cr}$-values.

The first-order (dashed-curves) and the corrected (solid-curve) amplitude variations with respect to plasma parameters are shown in Figs. 3 and 4 for compressive and rarefactive solitary profiles, respectively. The left/right columns of figures show the variation with respect to change in $\beta$(fixed $H$)/$H$(fixed $\beta$). The comparison of plots indicate that for values of $H$ very close to critical value (Fig. 3(d)) the correction is very large. Also it is remarked that for compressive excitations the total (corrected) amplitude increases with increase in the value of $\beta$ (Figs. 3(a)-3(c)), and conversely, the width of corrected profile tends to decrease. On the other hand, for rarefactive profiles the corrected amplitude tends to increase with increases of $\beta$-values (Figs. 4(a)-4(c)), while the width tends to increase.

\newpage

\textbf{FIGURE CAPTIONS}

\bigskip

Figure-1

\bigskip

(Color online) Dispersion curves of linear lower and higher electron-acoustic branches (Eq. (\ref{mode})) are shown for different values of fractional plasma degeneracy ($\beta$) and fixed value of $H$ (Fig. 1(a), 1(b)) and for different values of quantum diffraction parameter ($H$) and fixed value of $\beta$ (Fig. 1(c), 1(d)). Different dash sizes refer to and are related appropriately to different values of varied parameter in each plot.

\bigskip

Figure-2

\bigskip

(Color online) (a) A plot of $\beta$-$H$ plane showing the regions where the electron-acoustic solitary excitations are rarefactive (lower-region) or compressive (upper-region) and (b) Variation of the first-order soliton amplitude with respect to increase in the value of nondegenerate to degenerate electron equilibrium density for fixed other plasma parameters. (c) the variation of soliton width with respect to quantum diffraction parameter, $H$ showing critical values at which the width vanishes. Different dash sizes refer to and are related appropriately to different values of $\beta$ in Fig. 2(b) and the left-/right-branched in this plot represent where the solitons are compressive/rarefactive.

\bigskip

Figure-3

\bigskip

(Color online) Profiles for bright-type solitary excitations is given for first-order amplitude (dashed curves) and the corrected (total) amplitude (solid-curves) perturbations. The left/right columns of figures show the variation with respect to change in $\beta$(fixed $H$)/$H$(fixed $\beta$). The value of $M=-1$ is used for all plots in this figure.

\bigskip

Figure-4

\bigskip

(Color online) Profiles for dark-type solitary excitations is given for first-order amplitude (dashed curves) and the corrected (total) amplitude (solid-curves) perturbations. The left/right columns of figures show the variation with respect to change in $\beta$(fixed $H$)/$H$(fixed $\beta$). The value of $M=-1$ is used for all plots in this figure.

\newpage

\begin{figure}
\resizebox{1\columnwidth}{!}{\includegraphics{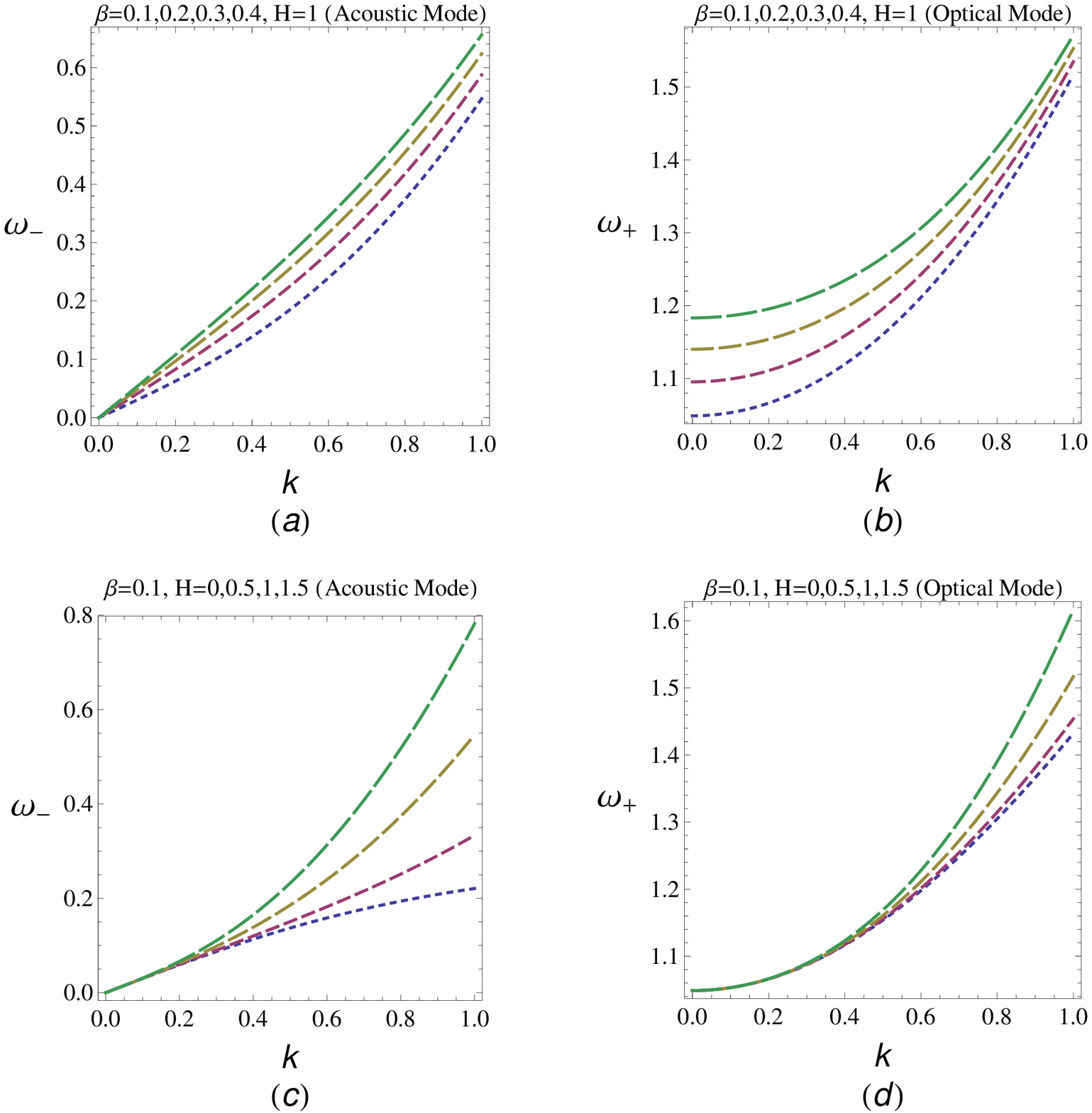}}
\caption{}
\label{fig:1}
\end{figure}

\newpage

\begin{figure}
\resizebox{1\columnwidth}{!}{\includegraphics{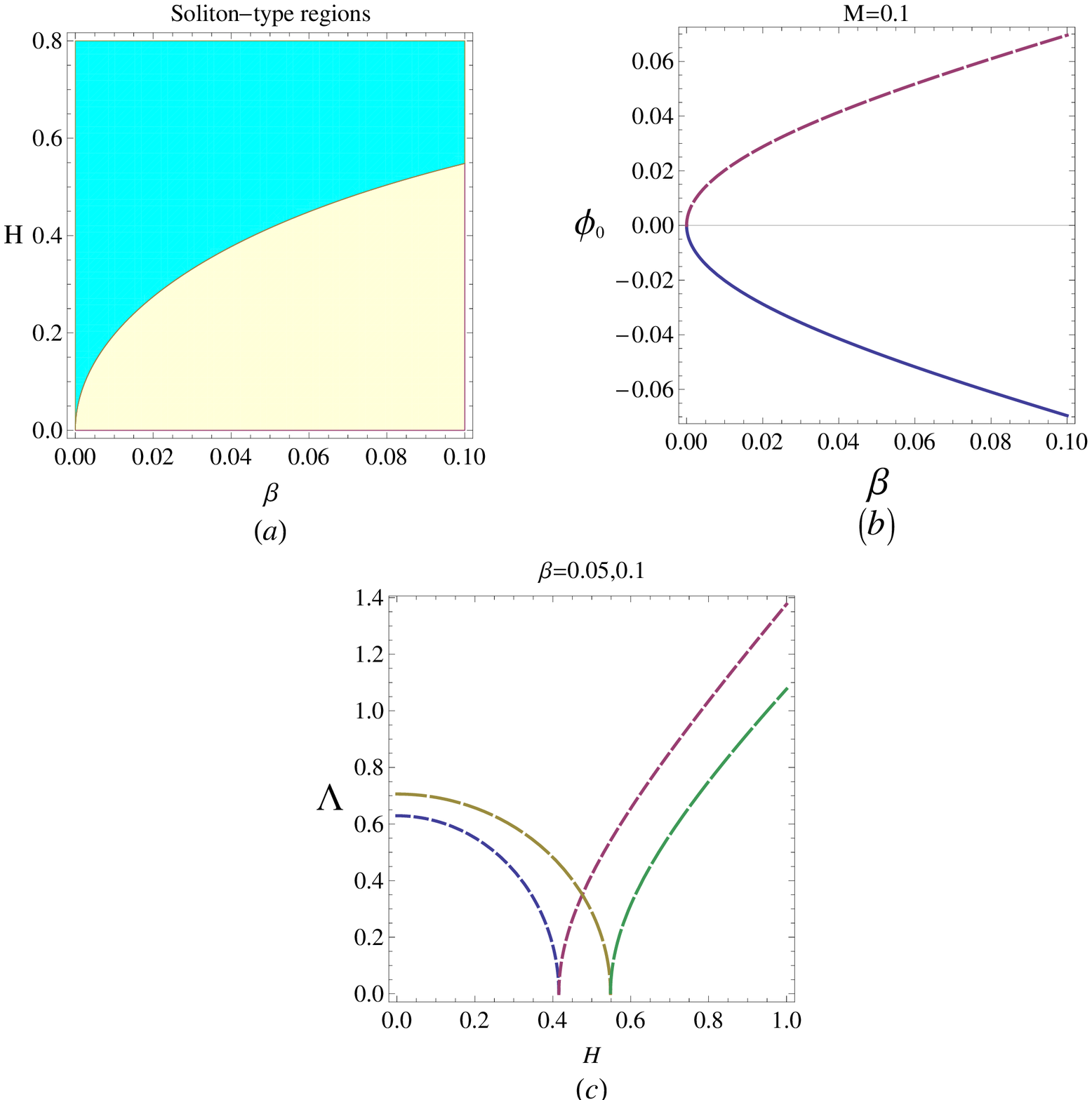}}
\caption{}
\label{fig:2}
\end{figure}

\newpage

\begin{figure}
\resizebox{1\columnwidth}{!}{\includegraphics{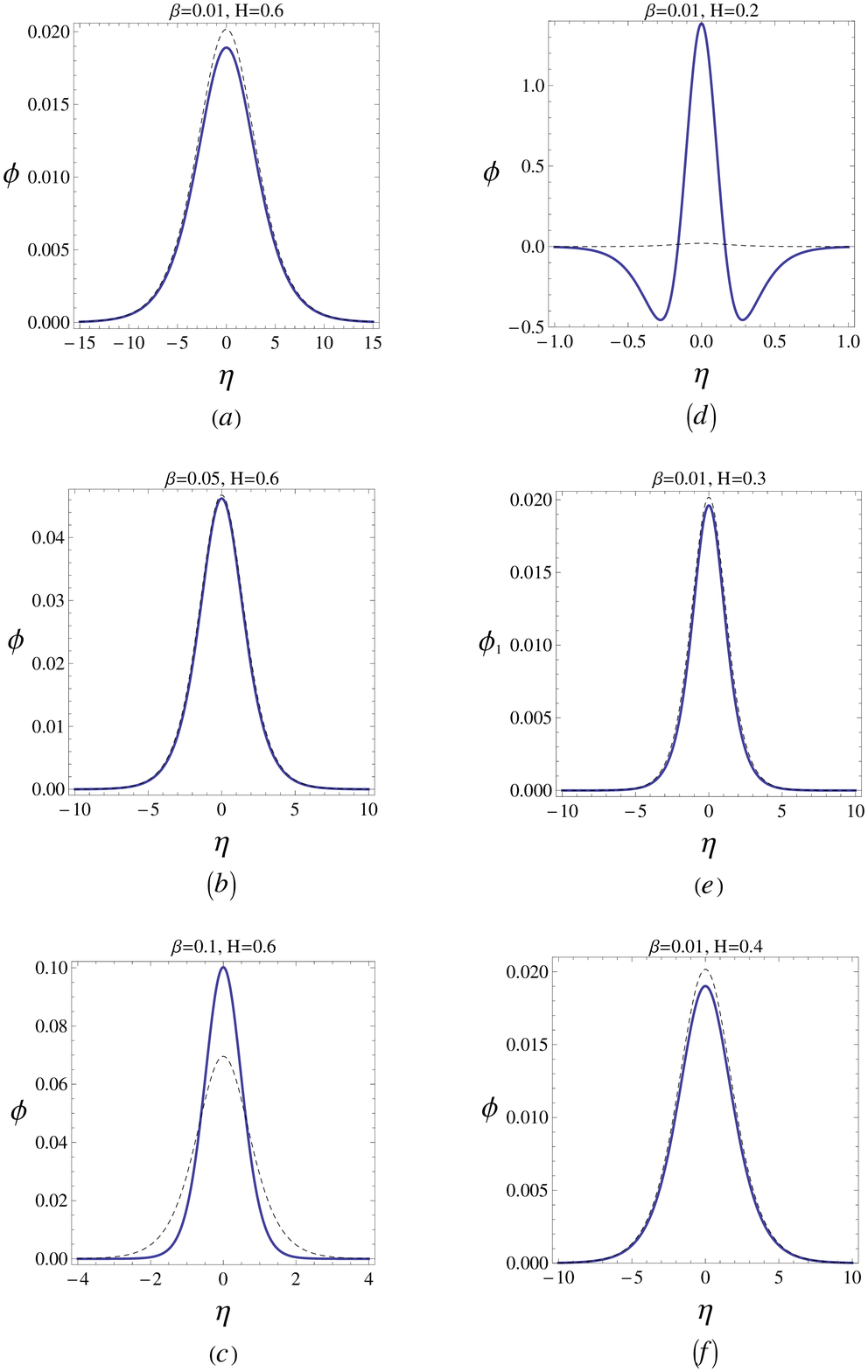}}
\caption{}
\label{fig:3}
\end{figure}

\newpage

\begin{figure}
\resizebox{1\columnwidth}{!}{\includegraphics{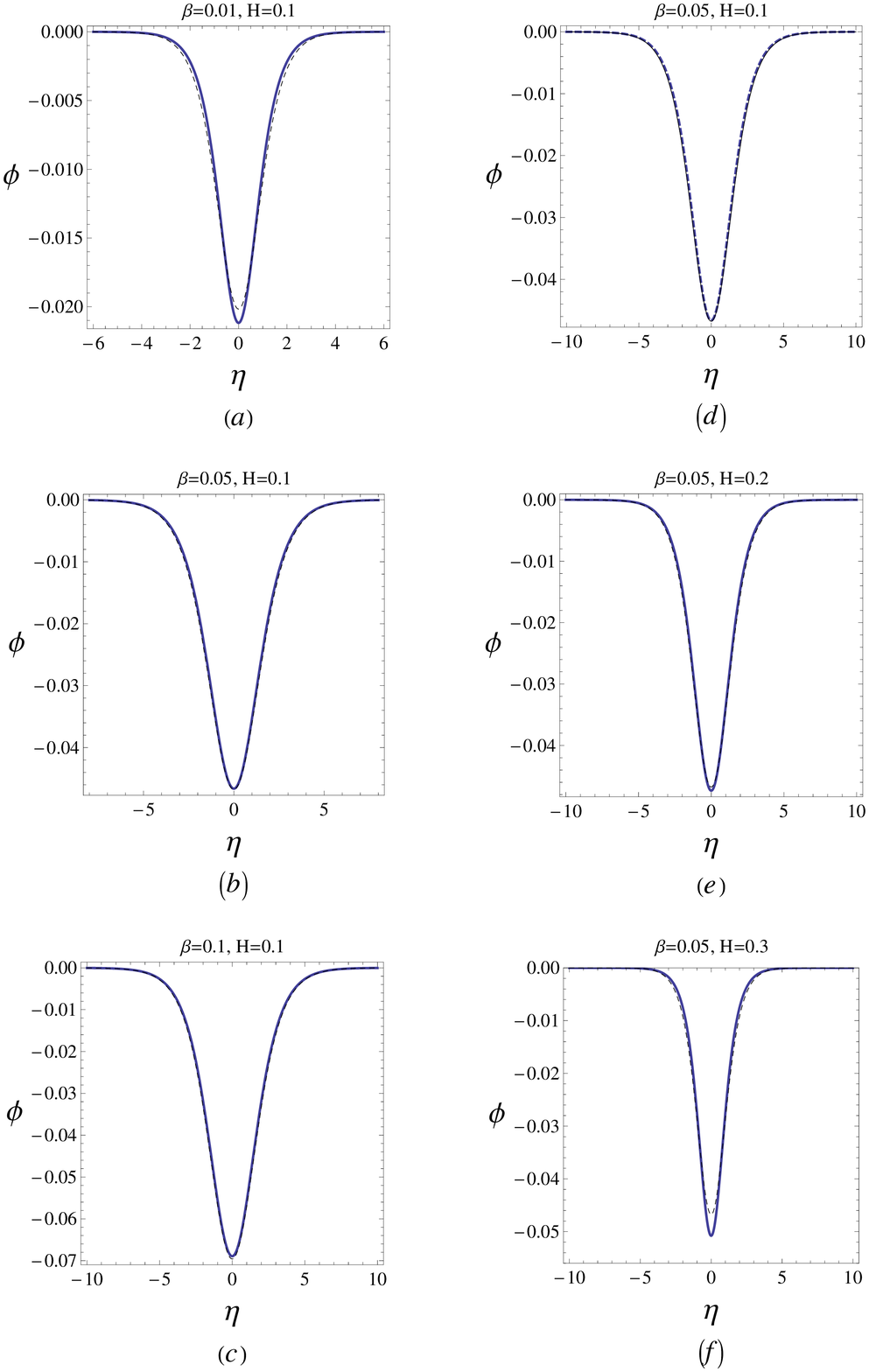}}
\caption{}
\label{fig:4}
\end{figure}

\end{document}